\documentstyle [12pt]{article}

\def\pa{\partial}

\begin{document}

\tolerance=5000

\begin{flushright}
OCHA-PP-60\\
NDA-FP-19\\
June (1995) \\
\end{flushright}

\vfill

\begin{center}
{\large\bf Is the Condensation of Strings \\the Origin of
Einstein Gravity ? }\\

       \vfill

{\sc Miyuki KATSUKI,\footnote{Present Adress: Fujitsu,
Kawasaki}
Hiroto KUBOTANI,
Shin'ichi NOJIRI${^\spadesuit}$,
Akio SUGAMOTO}\\
        \vspace{0.1in}

{\it Department of Physics, Faculty of Science\\
     Ochanomizu University \\
 1-1 Otsuka 2, Bunkyo-ku, Tokyo 112, JAPAN}\\
       \vspace{0.2in}
    ${^\spadesuit}${\it Department of
Mathematics and Physics\\
National Defence Academy\\
Yokosuka, 239, JAPAN}
\end{center}
\vfill
\begin{abstract}
A mechanism of generating the metric is proposed, where the
Kalb-Ramond symmetry existing in the topological BF theory
is broken through the condensation of the string fields
which are
so introduced as to couple with the anti-symmetric tensor
fields $B$, invariantly under the Kalb-Ramond symmetry.
In the chiral decomposition of the local Lorentz group,
the non-Abelian $B$ fields need to be generalized to
the string fields.
The mechanism of the condensation is discussed, viewing
the confinement problem and the polymer physics.
\end{abstract}
\newpage
\section{Introduction}
There is an interesting folklore (probably after
E. Witten \cite{Witten}) that at extremely high energy,
the concept of the metric disappears and the topological
nature of space-time only remains, where all the dynamical
degrees of freedom are washed away by the huge symmetries
existing in the theory called topological field theory.
When we go down to the lower energy, however, due to some
unknown phase transition, violation of the huge symmmeties
occurs and the usual Einstein gravity appears with the
recovery of the concept of the metric and of the local
degrees of freedom.

We know that the Einstein gravity can be
reformulated \cite{2-form gravity}, starting from a
topological theory (somebody calls BF theory \cite{B-F})
written in terms of the anti-symmetric tensor field
(2-form field) $B_{\mu\nu}^{AB}(x)$ and the Riemann tensor
$R_{\lambda\rho}^{AB}(x)=\pa_{\lambda}
\omega_{\rho}^{AB}(x)-\pa_{\rho}\omega_{\lambda}^{AB}(x)
+\omega_{\lambda}^{AC}(x)\omega_{\rho}^{CB}(x)
-\omega_{\rho}^{AC}(x)\omega_{\lambda}^{CB}(x)$,
given by the spin connection $\omega_{\mu}^{AB}(x)$ .
The action of this theory reads
\begin{equation}
S=\int\, \frac{1}{2} \epsilon^{\mu\nu\lambda\rho}
B_{\mu\nu}^{AB}(x)R_{\lambda\rho}^{AB}(x), \label{B-F}
\end{equation}
where we have used the Euclidean metric for the local Lorentz
frame, namely $A,\ B$ and $C$ take $1\sim4$.
If we fix the anti-symmetric tensor field in terms of the
vierbein $e_{\mu}^{A}$ as follows;
\begin{equation}
B_{\mu\nu}^{AB}(x) = \frac{1}{2} \epsilon^{ABCD}e_{\mu}^{C}
e_{\nu}^{D},
\label{the constraint}
\end{equation}
we have the Einstein gravity in the first order formalism.
The equation of motion for the spin connection,
$\nabla_{[\mu}^{AB}e_{\nu]}^{B} = 0$, can be solved in terms of
the vierbein, giving the usual Einstein gravity.
The original ``BF" action Eq.(\ref{B-F}) is invariant under the
local Lorentz group of $O(4)$, but it has the additional symmetry;
\begin{equation}
B_{\mu\nu}^{AB} \rightarrow B_{\mu\nu}^{AB}
+\nabla_{[\mu}^{AC}
\Lambda_{\nu]}^{CB},
\label{K-R}
\end{equation}
which is the non-Abelian version of the Kalb-Ramond symmetry
generated by the vector-like  parameter
$\Lambda_{\nu}^{BC}$ \cite{K-R}.
Owing to this additional symmetry the ``BF" theory becomes
topological one \cite{proof of the topological}.
Under the fixing of Eq.(\ref{the constraint}), the Kalb-Ramond
symmetry is broken and the Einstein gravity comes out.

Therefore, the understanding of the transition from the
topological theory to the Einstein gravity cannot be
realized without catching the physical meaning of the
constraint in Eq.(\ref{the constraint}).  Being not
satisfied with the transition  done by hand, we
are tempting to persue possible mechanisms of
generating the constraint dynamically.  This is
the motivation and the target of our paper.  In
this paper we will propose the condensation
mechanism of the extended objects (string fields)
in order to violate the Kalb-Ramond symmetry and
to generate the constraint condition.  Another
scenario is investigated in a separate paper \cite{Nojiri},
where the Kalb-Ramond symmetry is broken radiatively
to generate the metric.

\section{Chiral decomposition}

{}From the gauge theoretical viewpoint it is helpful to perform
the $SU(2)\times SU(2)$ decomposition of the $O(4)$ local Lorentz
group.  This makes the action Eq.(\ref{B-F}) into the following
form:
\begin{equation}
S=\int\,\frac{1}{2}
\epsilon^{\mu\nu\lambda\rho}
\left ( B_{\mu\nu}^{a}(x)R_{\lambda\rho}^{a}(x) +
\bar{ B}_{\mu\nu}^{a}(x)\bar{R}_{\lambda\rho}^{a}(x)\right).
\label{decomposed B-F}
\end{equation}
Here the first and the second terms represent, respectively,
the self-dual and anti-self-dual components;
\begin{eqnarray}
& & B_{\mu\nu}^{a}\equiv B_{\mu\nu}^{a4} + \frac{1}{2} \epsilon^{abc}
B_{\mu\nu}^{bc} \equiv \frac{1}{2}
\eta_{BC}^{a}B_{\mu\nu}^{BC} , \nonumber \\
& & \bar{B}_{\mu\nu}^{a}  \equiv
- \bar{B}_{\mu\nu}^{a4} +  \frac{1}{2}
\epsilon^{abc} \bar{B }_{\mu\nu}^{bc}  \equiv  \frac{1}{2}
\bar{\eta}_{BC}^{a}\bar{B}_{\mu\nu}^{BC}  , \\
& & R_{\mu\nu}^{a} = \pa_{[\mu}\omega_{\nu]}^{a}
+ \epsilon^{abc}
\omega_{\mu}^{b}\omega_{\nu}^{c}, \ \
\bar{R}_{\mu\nu}^{a} = \pa_{[\mu}\bar{\omega}_{\nu]}^{a}
+ \epsilon^{abc}\bar{\omega}_{\mu}^{b}\bar{\omega}_{\nu}^{c},
\nonumber
\label{self-dual decomposition}
\end{eqnarray}
with the corresponding $SU(2)\times SU(2)$
gauge fields
\begin{equation}
\label{gaugef}
\omega_{\mu}^{a} \equiv \frac{1}{2} \eta_{BC}^{a}
\omega_{\mu}^{BC},\ \
 \bar{\omega}_{\mu}^{a} \equiv
\frac{1}{2} \bar{\eta}_{BC}^{a}
\bar{\omega}_{\mu}^{BC},
\end{equation}
where $\eta_{BC}^{a}$ and $\bar{\eta}_{BC}^{a}$ denote
the 't Hooft
symbols \cite{'tHooft} for self-dual and anti-self-dual
decompositions, respectively, and  $a$, $b$ and $c = 1\sim 3$.
Corresponding to the  decomposition, the fixing in
Eq.(\ref{the constraint}) becomes
\begin{equation}
B_{\mu\nu}^{a} = \frac{1}{2} \eta_{BC}^{a} e_{\mu}^{B}
e_{\nu}^{C},\ \
\bar{B}_{\mu\nu}^{a} = \frac{1}{2} \bar{\eta}_{BC}^{a}
\bar{e}_{\mu}^{B}\bar{e}_{\nu}^{C},
\label{the decomposed constraint}
\end{equation}
which lead to the following relations;
\begin{eqnarray}
\label{C1}
\epsilon^{\mu\nu\lambda\rho}B_{\mu\nu}^{a}B_{\lambda\rho}^{b}
& = & 2e(x)\delta^{ab}, \\
\label{C2}
\epsilon^{\mu\nu\lambda\rho}\bar{B}_{\mu\nu}^{a}
\bar{B}_{\lambda\rho}^{b} & = & 2e(x)\delta^{ab},\\
\label{C3}
\epsilon^{\mu\nu\lambda\rho}B_{\mu\nu}^{a}
\bar{B}_{\lambda\rho}^{b} & = & 0 .
\end{eqnarray}
The relations in Eqs.(\ref{C1}) and (\ref{C2}) give the
constraints on the traceless symmetric part of the
indices $(a, b)$, which is the isospin-$2$ contribution
in the composition of the two isospin-$1$ components
$a$ and $b$, namely
\begin{eqnarray}
& & \epsilon^{\mu\nu\lambda\rho}\left( B_{\mu\nu}^{a}
B_{\lambda\rho}^{b} - \frac{1}{3} \delta^{ab}
B_{\mu\nu}^{c}B_{\lambda\rho}^{c}\right)=0,
\label{theC} \\
& & \epsilon^{\mu\nu\lambda\rho}\left( \bar{B}_{\mu\nu}^{a}
\bar{B}_{\lambda\rho}^{b} - \frac{1}{3} \delta^{ab}
\bar{B}_{\mu\nu}^{c}\bar{B}_{\lambda\rho}^{c}\right)=0.
\label{theC2}
\end{eqnarray}

The important point here is that if we impose the three
constraints in Eqs.(\ref{C3}), (\ref{theC}) and (\ref{theC2}),
then we can obtain  Eq.(\ref{the decomposed constraint}) as
the general solutions, by introducing an arbitrary
function $e_{\mu}^{A}(x)$ which will afterwards be
identified with the vierbein \cite{2-form gravity}.
If we restrict ourselves only to the self-dual
component of the anti-symmetric tensor fields
$B_{\mu\nu}^{AB}$ by setting  $\bar{B}_{\mu\nu}^{a} = 0$,
then we have only one constraint to be analyzed,
namely Eq.(\ref{theC}).  Therefore, If we can find
the dynamical origin of this constraint, we will surely
be very close to the dynamical generation of the metric
or the phase transition from the topological theory to
the Einstein gravity.  The same considertion can be
done for the anti-self dual component also.
Introducing a Lagrange multiplier field
$\phi^{ab}(x)$ with traceless isosin indices, we have
the following (chiral) action for the self-dual part:
\begin{equation}
  S = \int \frac{1}{2} \epsilon^{\mu\nu\lambda\rho}
\left( B^a_{\mu\nu}(x)R^a_{\lambda\rho}(x) +
\underbrace{ \phi^{ab}(x)B^a_{\mu\nu}(x)
B^b_{\lambda\rho}(x)}_{{\rm constraint\: term}} \right).
\label{chac}
\end{equation}

Now it is clearly recognized that without the constraint
term, the action Eq. (\ref{chac}) is invariant
under both (i) the $SU(2)$ local Lorentz transformation

\begin{eqnarray}
 \left\{
\matrix{
\omega^a_\mu & \rightarrow & \omega^a_\mu +
\epsilon^{abc}\omega^b_\mu \Lambda^c - \partial_\mu\Lambda^a \cr
B^a_{\mu\nu} & \rightarrow & B^a_{\mu\nu} +
\epsilon^{abc}B^a_{\mu\nu}\Lambda^c \cr
}
\right.
 \label{ SU(2) L-L }
\end{eqnarray}
and (ii) the Kalb-Ramond symmetry
\begin{eqnarray}
 \left\{
\matrix{
\omega^a_\mu & \rightarrow & \omega^a_\mu \cr
B^a_{\mu\nu} & \rightarrow & B^a_{\mu\nu} +
\nabla^{ab}_\mu \Lambda^b_\nu - \nabla^{ab}_\nu \Lambda^b_\mu \cr
}
\right.
\label{SU(2)K-R}
\end{eqnarray}
where  $\nabla^{ab}_\mu$ is the covariant derivative of
the $SU(2)$ gauge symmetry,
$ \: \nabla^{ab}_\mu\Lambda^b_\nu = \partial_\mu
\Lambda^a_\nu +
\epsilon^{abc}\omega^b_\mu\Lambda^c_\nu $.

A very naive estimation of the degrees of freedom of
$\omega_{\mu}^{a}$ and $B_{\mu\nu}^{a}$ in this theory is 30,
whereas the number of the independent gauge transformations
is 15 each of which kills two degrees of freedom, so that
the theory without the constraint (``BF theory") has no
local degrees of freedom, or is the topological theory.
The rigorous proof on this point can be found in
Ref.\cite{proof of the topological}.
The Kalb-Ramond symmety is, however, broken for
the action called ``2-form gravity" in
Eq.(\ref{chac}), where the constraint
Eq.(\ref{theC}) is added to the original ``BF" action.

\section{  String theory without the metric (Abelian case) }

We may remember that the Kalb-Ramond symmetry was originally
introduced as a gauge symmetry of the string
theory \cite{K-R}, where the anti-symmetric tensor
field $B^a_{\mu\nu}(x)$ plays the role of
the {\it gauge field} for the string  field
$\Psi[C]$ which is a functional defined on a
closed curve $C$.  In the Abelian case the
Kalb-Ramond transformation is given by
\begin{eqnarray}
\left\{
\matrix{
B_{\mu\nu} & \rightarrow & B_{\mu\nu} + \partial_\mu
\Lambda_\nu -
 \partial_\nu\Lambda_\mu
\label{AK-R1} \cr
\Psi[C] & \rightarrow & \exp ( i\oint _C dx^\mu
\Lambda_\mu(x) ) \Psi[C],
\label{AK-R2} \cr}
\right.
\end{eqnarray}
where $\Lambda_\mu(x)$ is the parameter of the
transformation.  If we define $\delta C^{\mu\nu}(x)$ to
be the rectangular deformation of a curve $C$ at $x$ in
the $\mu\nu$ direction, the derivative of the string
field can be defined as
\begin{equation}
    \frac{\delta}{ \delta C^{\mu\nu}(x) } \equiv
\lim_{{\rm area\: of\: }\delta
C^{\mu\nu}{\rm \to} 0 } \left(
\frac{ \Psi[C+\delta C^{\mu\nu}(x)] - \Psi[C] }
{{\rm area\: of\: }\delta C^{\mu\nu} }
\right)
\end{equation}
The covariant derivative is given by
\begin{equation}
  \frac{D}{DC^{\mu\nu}(x)} \equiv \frac{\delta}
{\delta C^{\mu\nu}(x)} - iB_{\mu\nu}(x),
\end{equation}
since it satisfies
\begin{eqnarray}
& & \left( \frac{\delta}{\delta C^{\mu\nu}(x)} -
iB{\rq}_{\mu\nu}(x) \right)
\Psi '[C] \nonumber \\
& & \hskip 1cm = \exp ( i\oint_Cdx^\mu\Lambda_\mu(x) )
\left( \frac{\delta}{\delta C^{\mu\nu}(x)} - iB_{\mu\nu}(x) \right)
\Psi[C]\ .
\end{eqnarray}
Now we have an action invariant under the Kalb-Ramond
transformation;
\begin{eqnarray}
  S &=& \int d^4 x\frac{1}{2} \epsilon^{\mu\nu\lambda\rho}
B_{\mu\nu}(x)R_{\lambda\rho}(x) \nonumber \\
& & + \sum_C \sum_{x( \in C )} \epsilon^{\mu\nu\lambda\rho}
\left[ \left( \frac{\delta}{\delta C^{\mu\nu}(x)} -
iB_{\mu\nu}(x) \right) \Psi[C] \right]^{\dagger}
\left[  \left( \frac{\delta}{\delta C^{\lambda\rho}(x)} -
iB_{\lambda\rho}(x) \right) \Psi[C]  \right] \nonumber \\
& & + \sum_{C} {\cal V } [ \Psi[C]^{\dagger} \Psi[C]]
\label{AK-Ra}
\end{eqnarray}

It is the characteristic point that this action is
defined using the $\epsilon$ symbol without using
the metric.  Therefore the theory may be a topological
one, where the configuration of the shape of strings may
be analyzed even in this theory with the $\epsilon$
symbol, but the vibration modes of the strings are
difficult to be measured without having the length or
the energy scale.  On the other hand, the string
theory defined using the metric  $g^{\mu [ \lambda}
g^{\nu\rho ] }$ in place of the
$\epsilon^{\mu\nu\lambda\rho}$ was studied more than
a decade ago as a field theory of ordinary vibrating
strings \cite{string}\cite{confinement}.

\section{Condensation of the strings and generation
of the metric (Abelian case)}

Now, we will consider the condensation of the strings
in the string theory given above.  Let us first
introduce  $\phi (x)$ as
\begin{equation}
\frac{1}{2} \phi (x) \equiv
\sum_{C (\owns x )} \Psi[C]^{\dagger} \Psi[C] .
\label{Abelian condensation}
\end{equation}
Then we have the following term in the action,
\begin{equation}
      S_{C} = \int d^4 x \frac{1}{2}
\epsilon^{\mu\nu\lambda\rho}
\phi(x) B_{\mu\nu}(x) B_{\lambda\rho}(x),
\label{S-C}
\end{equation}
which can be seen as the Abelian version of the
constraint term in the 2-form gravity,
Eq.(\ref{chac}).
Therefore, if the vacuum expectation value
$\langle \phi(x) \rangle $ of the field $\phi(x)$
is vanishing at high energies, then we have no
constraint on the anti-symmetric tensor field, that
is, we have the topological ``BF" theory, whereas
if it gives a large vacuum expectation value,
then we have the following constraint:
\begin{equation}
\frac{1}{2}\epsilon^{\mu\nu\lambda\rho} B_{\mu\nu}
B_{\lambda\rho} = 0 .
\end{equation}
The constraint can be viewed as the Abelian version
of the constraint in Eq.(\ref{theC}) which is necessary
to generate the concept of the metric and which
we are going to obtain dynamically.
The mechanism is a kind of the Meissner effect,
where the vacuum expectation value of the Higgs
field prohibits the penetration of the magnetic field;
\begin{equation}
       g^{\mu\nu} A_{\mu }A_{\nu} = 0 .
\end{equation}

The magnetic vortex condensation in the dual transformed
Higgs model induced the following term in the dually
transformed action \cite{confinement},

\begin{equation}
       -\frac{1}{4} (1 + \phi ) B^{\mu\nu}B_{\mu\nu},
\label {vortex condensation}
\end {equation}
which is just the term in Eq.(\ref{S-C}), if the
$g^{\mu [ \lambda} g^{\nu\rho ] }$ is used in the
place of $\epsilon^{\mu\nu\lambda\rho}$.
(Originally the notation $W_{\mu\nu}$ was used for
the present $B_{\mu\nu}$ \cite{K-R}\cite{confinement}.)
The generated term due to the vortex condensation was seen
in the original Higgs model as
\begin{equation}
       -\frac{1}{4} (1 + \phi )^{-1} F^{\mu\nu}F_{\mu\nu},
\label {confinement}
\end {equation}
which gives the anti-Meissner effect or the confinement
when the condensation of the string field occurs having
the large value of the $\phi$, where the $F_{\mu\nu}$
is the field strength of the electromagnetic field.

\section{String theory without the metric (non-Abelian case)}

In order to understand the gravity, we should generalize
the above discussion starting from
Eq.(\ref{AK-Ra}) to the non-Abelian case.
When we write down the  non-Abelian version of the
transformation in Eq.(\ref{AK-R2}), we should be
careful about the line integral of the transformation
parameter $\Lambda_{\mu}^{a} (x)$, since it has
different transformation properties at different positions.
In order to have the definite transformation property
under the local $SU(2)$, we should make all the
points on the curve $C$ be connected to a
special point $x_0$ by the Wilson operator
and the different transformation properties
at different points be unified to a single
one at $x_0$.  Then, we have the following
non-Abelian version of the Kalb-Ramond
transformation,
\begin{equation}
\label{nAK-R1}
\Psi^i[C;x_0]  \rightarrow
U[C;x_0]^i_j\Psi^j[C;x_0]
\end{equation}
with
\begin{equation}
U[C;x_0]^i_j \equiv \left[ \exp \left\{
 i \oint_C dx^{\mu} \Lambda^a_\mu(x) W[ x\leftarrow x_0]^a_b
 T^b_R \right\} \right]^i_j
\label{nAK-R2}
\end{equation}
where $\Psi[C; x_0]$ behaves as a local field at
$x_0$ under the local $SU(2)$ transformation
in the representation $\{ R\}$ and the Wilson
operator is defined using the adjoint representation $\{A\}$ ;
\begin{equation}
   W_P [x\leftarrow x_0]^a_b = \left[ P \: \exp i
\int_{P( x\leftarrow x_0)} dx^\mu ( \Lambda_\mu^c T_A^c )
\right]^a_b ,
 \label{ Wilson operator}
\end{equation}
Here the special point $x_0$ and the path
$P( x\leftarrow x_0)$ are chosen to be along
the curve $C$. (It is also possible to take $x_0$
outside the curve $C$ and the path $P$ not along the
curve $C$, but the above choice is the simpler one.)
Corresponding to the non-Abelian Kalb-Ramond
transformation for the ``matter string fields" in
Eq.(\ref {nAK-R1}), we can find the following
transformation-property for the ``gauge string fields"
$B_{\mu\nu}[C; x, x_0]$ in the matrix notation, by
generalizing the non-Abelian Kalb-Ramond transformation
for $B_{\mu\nu}(x)$ in Eq.(\ref{SU(2)K-R}), that
is,

\begin{equation}
B_{\mu\nu}[C; x, x_0] \rightarrow U[C;x_0]
B_{\mu\nu}[C; x, x_0] U[C;x_0]^{-1}
- \frac{\delta U[C;x_0]}{\delta C^{\mu\nu}} U[C;x_0]^{-1}.
\label{nAK-R3}
\end{equation}
Both equations (\ref{nAK-R1}) and
(\ref{nAK-R3}) gurantees the existence
of the covariant derivative
\begin{equation}
\label{covariant derivative for the string}
\frac{D}{DC^{\mu\nu}} \equiv \frac{\delta}{\delta C^{\mu\nu}}
+ B_{\mu\nu}[C; x, x_0]
\end{equation}
satisfying
\begin{equation}
\frac{D}{DC^{\mu\nu}}\Psi[C;x_0] \rightarrow U[C;x_0]
\frac{D}{DC^{\mu\nu}}\Psi[C;x_0] .
\end{equation}
The transformation of the Kalb-Ramond string field
$B_{\mu\nu} [C; x, x_0]$ in Eq.(\ref{nAK-R3})
can be written more explicitly in terms of the
components $B^a_{\mu\nu} [C; x, x_0]$,
\begin{eqnarray}
& & B^a_{\mu\nu}[C; x, x_0] \rightarrow \nonumber \\
& & \hskip 0.5cm B^a_{\mu\nu}[C; x, x_0]
 - \epsilon^{abc}\oint_C dy^{\nu} \Lambda^{b'}_{\nu}(y)
W_C[y\leftarrow x_0]_{b'b} B^c_{\mu\nu}[C; x, x_0]
\nonumber \\
& & \hskip 0.5cm - i \Bigl\{
(\nabla_{\mu} \Lambda_{\nu}(x) - \nabla_{\nu}
\Lambda_{\mu}(x))^{a'} W_C[x\leftarrow x_0]_{a'}^{a}
\nonumber \\
& & \hskip 0.7cm + \int_{y\ge x} dy^{\nu} \Lambda^b_{\nu}(y)
W_C[y\leftarrow x]_{bb'} (i T^c_A)^{b'}_{a'}
R^c_{\mu\nu}(x) W_C[x\leftarrow x_0]^{a'a}
\Bigr\}.
\label{nAK-Re}
\end{eqnarray}
This transformation property is rather different from
the original non-Abelian version of the Kalb-Ramond
transformation in Eq.(\ref{SU(2)K-R}), but if the
closed curve $C$ becomes very small, the transformation
approaches to
\begin{equation}
B^a_{\mu\nu}[C; x, x_0] \rightarrow
B^a_{\mu\nu}[C; x, x_0] - i (\nabla_{\mu}\Lambda_{\nu}
- \nabla_{\nu}\Lambda_{\mu} )^a,
\label{sCK-R}
\end{equation}
so that in the small $C$ limit
we have recover the original
Kalb-Ramond field;
\begin{equation}
B^a_{\mu\nu}(x) = \lim_{C\rightarrow {\rm small}}
(-i) B^a_{\mu\nu}[C; x, x_0] .
\label{defK-Rf}
\end{equation}

Now, we can write down the invariant action under
the non-Abelian version of the Kalb-Ramond
transformations obtained above in
Eqs.(\ref{nAK-R1}),
(\ref{nAK-R2}), (\ref{nAK-R3}) and (\ref{nAK-Re}):
\begin{eqnarray}
  S &=& \int d^4 x\frac{1}{2} \epsilon^{\mu\nu\lambda\rho}
B^a_{\mu\nu}(x)R^a_{\lambda\rho}(x) \nonumber \\
& & + \sum_C \sum_{x_0 (\in C)} \sum_{x( \in C )}
\epsilon^{\mu\nu\lambda\rho}
\left[ \left( \frac{\delta}{\delta C^{\mu\nu}(x)} +
T^a B^a_{\mu\nu}[C; x, x_0]\right) \Psi[C; x_0] \right]
 ^{\dagger} \nonumber \\
& & \hskip 1cm \times \left[  \left( \frac{\delta}
{\delta C^{\lambda\rho}(x)} +
T^a B^a_{\lambda\rho}[C; x, x_0]\right) \Psi[C; x_0]
\right] \nonumber \\
& & + \sum_{C}\sum_{ x_0} {\cal V }
 [ \Psi[C; x_0] ^{\dagger} \Psi[C; x_0]] .
\label{nAK-Ra}
\end{eqnarray}

\section{ Condensation of the strings and generation
of the metric (non-Abelian case) }

Having the non-Abelian action given above, if we will
define $\phi^{ab}[C; x_0]$ by
\begin{equation}
 \frac{1}{2}\phi^{ab}[C; x_0] \equiv
\Psi[C; x_0] ^{\dagger} T^a_R T^b_R\Psi[C; x_0] ,
\label{non-Abelian condensation}
\end{equation}
then we have the following contribution in the action, namely
\begin{equation}
S_{C}' = \sum_{C}\sum_{x_0 (\in C)} \sum_{x( \in C )}
\frac{1}{2}    \epsilon^{\mu\nu\lambda\rho}
\phi^{ab}[C; x_0] B^a_{\mu\nu}[C; x, x_0]
B^b_{\lambda\rho}[C; x, x_0] .
\label{S-Cprime}
\end{equation}

Now we will give the same discussion as in
the Abelian case on the condensation of the
string field: If the vacuum expectation value
$\langle \phi^{ab}[C; x_0] \rangle$ of the
isospin-$2$ part (traceless part with respect
to $(a,b)$) of the $\phi^{ab}[C; x_0]$ is
vanishing even in the small $C$ limit, then
we have no constraint given in Eq.(\ref{theC}),
so that we are still in the topological phase.
If the small $C$ limit of the isospin $2$ part
of the condensation
$\langle \phi^{ab}[C; x_0]\rangle$ gives,
fortunately, large non-vanishing values,
then we obtain the required constraint
in Eq.(\ref{theC}).  Therefore the
condensation of the string field, the ``matter"
for the ``Kalb-Ramond gauge field" $B^a_{\mu\nu}(x)$,
may trigger the phase transition from the
topological ``BF" theory to the 2-form gravity
which is the chiral part of the Einstein gravity
and in which the concept of the metric is recovered.

\section{ Discussion with the quark confinement and
the polymer physics}

We have as yet not enough ability to clarify the dynamics
for the above mentioned condensation of the string fields
to ocuur.  It is, however, the mechanism itself is very
similar to the proof of the quark confinement as was
discussed above in the Abelian case \cite{confinement}.
Both mechanisms of the generation of the metric and of
the quark confinement are formally interchanged, according
to the replacement of the roles of the epsilon symbol
$\epsilon^{\mu\nu\lambda\rho}$ and the product of the
metric, $g^{\mu [\lambda}g^{\nu\rho]}$. But, the
condensation of the string field may commonly
underlie the both phenomena.  The Coleman-Weinberg
mechanism performed before \cite{confinement}
in the study of the quark confinement due to
the vortex condensation might be helpful.
The importance of the radiative corrections
on our problem can also be found in Ref.\cite{Nojiri}
and Ref.\cite{radiative correction}.
In the polymer physics (a kind of string theory with
some interactions), a similar condensation mechanism of
the spin 2 field exists \cite{polymer}.  If we define
${\bf u}$ as a unit vector representing the direction
of the monomer, then the expectation value
$\langle u^a u^b \rangle$ ($a = 1\sim 3$) is non-vanishing
when the stress, given by the stress tensor $\sigma^{ab}$,
is applied from outside to the polymer solution: We have
\begin{eqnarray}
   Q^{ab} &\equiv& \langle (u^a u^b - \frac{1}{3} \delta^{ab})
\rangle \cr
&\propto& (\sigma^{ab}
- \frac{1}{3} \delta^{ab}\sigma^c_c) \cr
&\propto& (\varepsilon^{ab} - \frac{1}{3}
\delta^{ab}\varepsilon^c_c),
\label{polymer}
\end{eqnarray}
where $Q^{ab}$ is called ``directional order parameter"
and $\varepsilon^{ab}$ is the electric permeability
of the polymer solution.  The last equation gives the
optoelasticity, which can be undersood as
{\it the generation of the metric or of the
distortion of the space} causing the double
refraction, or the gravitational lensing effect
in our terminology.  The dynamical origin of
such an interesting phenomena is
{\it the entropy effects} (the effect of
summing up all the possible shapes of polymers)
as well as {\it the interactions between the monomers}
such as the nematic interaction of the liquid crystals.
Therefore, in order to understand the gravity,
especially the generation of the metric, we
should persue the condensation mechanism of
the extended objects (the string fields)
based on the entropy effects as well as
the interactions between the portions of
the strings (the monomers), following
the polymer dynamics.



\begin{thebibliography}{99}
\bibitem{Witten}  E. Witten, {\sl Nucl. Phys.} {\bf B 311}
(1988/1989) 46.
\bibitem{2-form gravity} J.F. Plebanski,
{\sl J. Math. Phys.} {\bf 18 } (1977) 2511;\\
A. Ashtekar, {\sl Phys. Rev. Lett.} {\bf 57 } (1986) 2244;
{\sl Phys. Rev.} {\bf D36 } (1987) 1587; \\
R. Capovila, T. Jacobson and J. Dell,
{\sl Phys. Rev. Lett.} {\bf 63 } (1989) 2325;
R. Capovila, J. Dell,  T. Jacobson and M. Mason,
{\sl Class. Quantum Grav.} {\bf 8 } (1991) 41;\\
G. 't Hooft, {\sl Nucl. Phys.} {\bf B357 } (1991) 211;\\
See also the good review article: H. Ikemori, in
{\it Proceedings of the Workshop on Quantum
Gravity and Topology }, ed. by I. Oda
( INS-Report No. 
\bibitem{B-F} A.S. Schwarz, {\sl Commun. Math. Phys.}
{\bf 67} (1979) 1;\\
G.T. Horowitz, {\it i.b.d. }{\bf 125} (1989) 417;\\
M. Blau and G. Thompson, {\sl Ann. Phys.}
{\bf 125 } (1991) 130; \\
I. Oda and S. Yahikozawa, {\sl Phys. Lett.}
{\bf B234 } (1990) 69;
{\bf B238 } (1990) 272; {\sl Prog. Theor. Phys.}
{\bf 83 } (1990) 845.
\bibitem{K-R} M. Kalb and P. Ramond, {\sl Phys. Rev.}
{\bf D9 }
(1974) 2273;\\
Y. Nambu, in {\it Quark Confinement and Field Theory },
Proceedings of the Rochestor Conference (1976) ed. by
D.R. Stump and D.H. Weingarten (Wiley, New York, 1977);
{\sl Phys. Rep.} {\bf 23C } (1976) 250;\\
A. Sugmoto, Phys. Rev. {\bf D19 } (1979) 1820; K. Seo,
M. Okawa and A. Sugamoto, {\sl Phys. Rev.}
{\bf D19 } (1979) 3744;
K. Seo and M. Okawa, {\sl Phys. Rev.}
{\bf D21 } (1980) 1614; \\
D.Z. Freedman and P.K. Townsend, {\sl Nucl. Phys.}
{\bf B177 }
(1981) 282.\\
See also the review article on the dual
transformation: R. Savit, {\sl Rev. Mod. Phys.} {\bf 52 }
(1980) 453.\\
\bibitem{proof of the topological}
M. Blau and G. Thompson, {\sl Phys. Lett.}
{\bf B255 } (1991) 535;\\
H.Y. Lee, A. Nakamichi and T. Ueno, {\sl Phys. Rev.}
{\bf D47} (1993) 1563.
\bibitem{Nojiri} M. Katsuki, S. Nojiri and
A. Sugamoto, \lq\lq Two-Form Gravity and the
Generation of Space-Time'', preprint OCHA-PP-
and NDA-FP-    June (1995) .
\bibitem{radiative correction}
A. Nakamichi, PhD thesis, \lq\lq {\it Quantum Corrections
in Topological 2-form Gravity }''
(Tokyo Institute of Technology, 1994);\\
I. Oda and S. Yahikozawa, {\sl Class. Quantum Grav.}
{\bf 11 } (1994) 2653.
\bibitem{string}
C. Marshall and P. Ramond, {\sl Nucl. Phys.}
{\bf B85 } (1975) 375.
\bibitem{confinement}
K. Seo and A. Sugamoto, {\sl Phys. Rev.}
{\bf D24 } (1981) 1630.
\bibitem{polymer}
M. Doi, in {\it Polymer Physics and Phase
Transition Dynamics}, Modern Physics Series No.19
(Iwanami Pub., 1992 ) (in Japanese); M. Doi and S. F. Edwards,
\lq\lq {\it The Theory of Polymer Dynamics}''
(Oxford Univ. Press, 1986 ) .
\bibitem{'tHooft}G. 't Hooft, {\sl Phys. Rev.} {\bf D14}
(1976) 3432.
\end{thebibliography}
\end{document}